\documentclass[english,pra,lengthcheck,final,superscriptaddress]{revtex4-1}
\usepackage[T1]{fontenc}
\usepackage[utf8]{inputenc}
\usepackage{color}
\usepackage{babel}
\usepackage{amsmath}
\usepackage{amssymb}
\usepackage{graphicx}
\usepackage{wasysym}
\usepackage[unicode=true,pdfusetitle,
 bookmarks=true,bookmarksnumbered=false,bookmarksopen=false,
 breaklinks=false,pdfborder={0 0 1},backref=false,colorlinks=true]
 {hyperref}

\makeatletter
\usepackage{physics}

\makeatother

\begin{document}
\title{Maximizing information obtainable by quantum sensors through the Quantum
Zeno Effect}
\author{Bruno Ronchi}
\affiliation{Instituto Balseiro, CNEA, Universidad Nacional de Cuyo, S. C. de Bariloche,
Argentina.}
\author{Analia Zwick}
\affiliation{Instituto Balseiro, CNEA, Universidad Nacional de Cuyo, S. C. de Bariloche,
Argentina.}
\affiliation{Centro At{\'o}mico Bariloche, CONICET, CNEA, S. C. de Bariloche,
Argentina.}
\affiliation{Instituto de Nanociencia y Nanotecnologia, CNEA, CONICET, S. C. de
Bariloche, Argentina}
\author{Gonzalo A. {\'A}lvarez}
\email{Corresponding author: gonzalo.alvarez@conicet.gov.ar}

\affiliation{Instituto Balseiro, CNEA, Universidad Nacional de Cuyo, S. C. de Bariloche,
Argentina.}
\affiliation{Centro At{\'o}mico Bariloche, CONICET, CNEA, S. C. de Bariloche,
Argentina.}
\affiliation{Instituto de Nanociencia y Nanotecnologia, CNEA, CONICET, S. C. de
Bariloche, Argentina}
\begin{abstract}
Efficient quantum sensing technologies rely on precise control of
quantum sensors, particularly two-level systems or qubits, to optimize
estimation processes. We here exploit the Quantum Zeno Effect (QZE)
as a tool for maximizing information obtainable by quantum sensors,
with a specific focus on the level avoided crossing (LAC) phenomenon
in qubit systems. While the estimation of the LAC energy splitting
has been extensively studied, we emphasize the crucial role that the
QZE can play in estimating the coupling strength. We introduce the
concept of information amplification by the QZE for a LAC system under
off-resonant conditions. The proposed approach has implications for
AC magnetic field sensing and the caracterization of complex systems,
including many-spin systems requiring the estimation of spin-spin
couplings. Overall, our findings contribute to the advancement of
quantum sensing by leveraging the QZE for improved control and information
extraction.
\end{abstract}
\maketitle

\section{Introduction}

Developing efficient quantum sensing technologies at atomic and nanometric
scales is critically dependent on precise control of the sensors to
enable the achievement of optimal estimation processes \citep{Meriles2010,Almog2011,Bylander2011,Alvarez2011,Smith2012,alvarez_controlling_2012,Cywinski2014,suter_colloquium:_2016,zwick16,zwick16crit,Degen2017,Poggiali2018,zwick2020precision,zwick2023quantum}.
The simplest quantum sensor is a two-level system –a qubit– and a
variety of experimental desings have been proposed for this purpose
\citep{Degen2017}, including cold atoms \citep{Duan2003,Bakr2009,Diehl2008},
ion traps \citep{Jurcevic2014,Martinez2016,Landsman2019}, Rydberg
atoms \citep{Saffman2010,Schachenmayer2013}, polar molecules \citep{Yan2013,Schachenmayer2013},
nuclear spins in liquids and solids \citep{Zhang2008,Alvarez2011,Smith2012,Alvarez2010,Souza2011,Alvarez2013,Alvarez2015,Yang2020,Jiang2021},
and nitrogen-vacancy centers (NVc) in diamonds \citep{Fischer2013,Staudacher2013,Shi2013,Waldherr2014,alvarez_local_2015,Staudacher2015,schmitt2017submillihertz,boss2017quantum,glenn2018high,Zangara2019,jiang2023quantum,segawa2023nanoscale}.

A two-level system can be a subset of many energy levels from a more
complex system. Most physical interactions involving two-level systems
results in a phenomenon known as a level avoided crossing (LAC) \citep{landau2013quantum}.
This happens when the energy levels of the two-level system approach
each other without actually crossing, as a result of interactions.
The most general Hamiltonian describing a two-level system includes
a coupling interaction between the energy levels, facilitating transitions
between them, and a ``longitudinal'' energy offset that creates
the energy splitting between the levels.

When employing a two-level system for quantum sensing, it is essential
to estimate the relevant interactions, such as the energy offset and/or
the coupling between the states \citep{zwick16,zwick2023quantum,Degen2017}.
Accurately characterizing the corresponding LAC Hamiltonian is thus
crucial for various applications, including improving spectroscopy
techniques \citep{PhysRevX2015.5.041016,Smith2012}, hyperpolarization
in NMR \citep{pravdivtsev2013exploiting,pravdivtsev2014exploiting,ivanov2014role,pravdivtsev2014highly},
NV center sensors \citep{king2015room,suter_rao2020level,wang2013sensitive,alvarez_local_2015,broadway2016anticrossing,ajoy2018orientation,Zangara2019}
and in general for developing quantum devices that operate at avoided
crossings \citep{PRXQuantum2021.2.010311}. 

While the energy splitting of a two-level system has been extensively
studied, and optimal methods for estimating it in different setups
have been explored \citep{pang2017optimal,Degen2017}, the optimization
of the estimation procedure for determining the coupling strength
between the two energy levels still requires further explorations
\citep{PhysRevLett.113.210404,Molmer_2015,joas2021online}. The coupling
strength varies due to the qubit's environment and serves as a crucial
source of information in several applications, including characterizing
spin-spin coupling network topologies \citep{alvarez_controlling_2012}
and time-dependent magnetic fields \citep{Degen2017,ciurana2017entanglement,Poggiali2018,mishra2022integrable,kiang2024quantum,Kuffer2022a}.

Quantum estimation tools can provide optimal strategies for efficiently
maximizing and extracting information about these relevant parameters
\citep{Caves_1994_fisher,Paris2009_QUANTUM-ESTIMATION,escher2011general,Paris2014_Characterization-of-classical-Gaussian,zwick16crit,Degen2017,Zwick2019_mukherjee_enhanced,zwick2020precision}.
Depending on the specific parameters, different strategies may be
most effective \citep{zwick16}. For example, coherent control may
be the optimal strategy for estimating some parameters, while incoherent
control techniques such as stroboscopic measurements \citep{kwiat1999high,kofman00,alvarez_controlling_2012,zwick16,muller2016stochastic,muller2020noise}
capable of slowing down the decoherence process, know as the quantum
Zeno effect (QZE) \citep{misra77,kofman00}, may be preferable for
estimating other parameters \citep{alvarez_controlling_2012,Molmer_2015,zwick16,Do2019,muller2020noise,Virzi2022}.

The QZE can be use to control the transfer of magnetization between
spins and significantly amplify the signal emitted by the spins \citep{kurizki_resonant_1996,kofman00,erez2008thermodynamic,Alvarez2010,alvarez_controlling_2012,dasari2021anti}.
This manipulation has allowed experimental determination of interactions
between spins in networks of many interacting spins \citep{alvarez_controlling_2012}.
By inducing the QZE through stroboscopic measurements, the system
dynamic is simplified from a complex behaviour depending on several
parameters to a simpler one based on a smaller number of parameters.
This simplification allows for a more direct determination of coupling
strengths. Complementing these results, quantum information metrics
have shown that a quantum probe can more efficiently determine the
coupling with its unknown surrounding environment when its dynamic
is steered by the QZE \citep{zwick16}.

In this work, we explore when stroboscopic measurements can enhance
the information extraction about the coupling determined by the level
avoided crossing. While previous studies suggested limited utility
of the QZE for estimating the coupling between states in a LAC Hamiltonian
under resonant condition \citep{Molmer_2015}, our findings demonstrate
that incoherent control, such as stroboscopic measurements, can be
particularly beneficial when the system is off-resonance. We introduce
the concept of information amplification by the QZE in a qubit sensor
with an offset from resonance, using quantum information tools. This
concept includes key elements relevant to more complex systems, such
as many-spin systems requiring the estimation of spin-spin couplings
for Hamiltonian and/or molecular characterization \citep{alvarez_controlling_2012}.

\section{\label{sec:Hamiltonian-parameter-estimation}Coupling strength estimation
of a qubit-probe}

The general Hamiltonian 
\begin{align}
H=\frac{1}{2} & \pmb\omega\cdot\pmb\sigma,\label{eq:Hamiltonian}
\end{align}
describes a two-level spin system acting as a qubit-probe, where the
vector $\pmb\sigma=(\sigma_{x},\sigma_{y},\sigma_{z})$ contains the
Pauli spin operators, and $\pmb\omega=(\omega_{x},\omega_{y},\omega_{z})$
the spin precesion frequency. The intrinsic energy splitting between
the spin states $\left|\uparrow\right\rangle $ and $\left|\downarrow\right\rangle $
is denoted by $\omega_{z}$ considering $\hbar=1$. For simplicity,
we assume $\omega_{y}=0$ so as the component $\omega_{x}$ gives
the coupling strength between the two qubit states $\left|\uparrow\right\rangle $
and $\left|\downarrow\right\rangle $. This Hamiltonian provides a
universal general form associated with a coupled two-level system,
giving rise to a LAC as a function of the energy splitting $\omega_{z}$,
due to the coupling $\omega_{x}$. The system is on resonance at $\omega_{z}=0$,
as the population exchange between the states $\left|\uparrow\right\rangle $
and $\left|\downarrow\right\rangle $ can occur completely. When $\omega_{z}\ne0$,
it gives the offset of the resonance condition, and the population
exchange probability between the states $\left|\uparrow\right\rangle $
and $\left|\downarrow\right\rangle $ is attenuated by increasing
$\omega_{z}$.

The density matrix for the two-level system state is
\begin{align}
\rho=\frac{1}{2}\left[\mathrm{I}+\pmb\mu\cdot\pmb\sigma\right] & ,\label{eq:rho}
\end{align}
where $\pmb\mu$ is the polarization vector. Initializing the qubit-probe
with a given polarization on the $\mathbf{z}$ direction $\pmb\mu(0)=\mu_{0}\mathbf{z}$
at time $t=0$, the qubit state evolves coherently as a function of
time $t$, described by a precesion of the polarization vector $\pmb\mu$
with the angular frequency $\pmb\omega$. Then, the observable is
the polarization along the longitudinal axis $\left\langle \sigma_{z}\right\rangle $,
which evolves over time depending on the spin precesion frecuency
$\pmb\omega$.

By measuring the observable $\left\langle \sigma_{z}\right\rangle $,
the information of the coupling strength $\omega_{x}$ between the
two levels $\left|\uparrow\right\rangle $ and $\left|\downarrow\right\rangle $
is encoded on the evolution of the qubit-state probablities $p^{\uparrow}(t)=\frac{1+\mathbf{z}\cdot\pmb\mu}{2}$
and $p^{\downarrow}(t)=\frac{1-\mathbf{\mathbf{z}}\cdot\pmb\mu}{2}$
to obtain $\left|\uparrow\right\rangle $ and $\left|\downarrow\right\rangle $
respectively. The information about the parameter is encoded differently
if the qubit sensor undergoes coherent or incoherent dynamics. To
determine and quantify the best strategy for estimating the coupling
strentgh $\omega_{x}$, depending on the offset $\omega_{z}$ and
the available experimental time, we resort to quantum estimation tools
\citep{Caves_1994_fisher,Paris2009_QUANTUM-ESTIMATION,escher2011general,amari_2016}.

\subsection{Fisher information of the coupling strength}

\begin{figure}
\centering

\includegraphics[width=0.95\columnwidth]{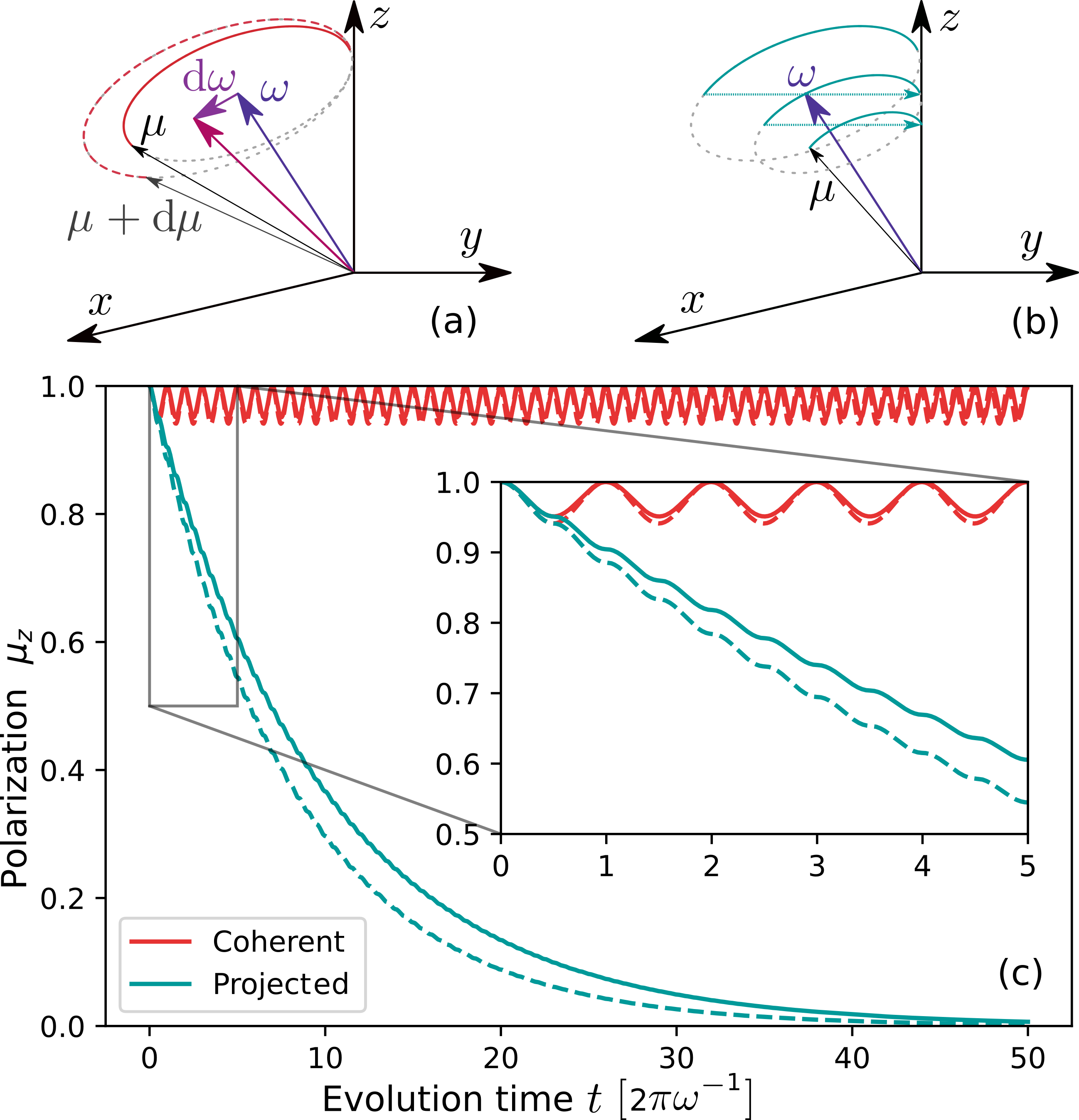}

\caption{Comparison of the qubit-probe trajectories of the polarization vector
$\pmb\mu$ for coherent and projected evolutions for the same precesion
frequency $\pmb\omega$. (a,b) Polarization vector $\pmb\mu$ (black
vector) for coherent (solid red line) and projected evolution (solid
blue line) for the same precesion frequency $\pmb\omega$ (violet
vector). The dotted lines show the continuation of the polarization
trajectory. In the coherent evolution (a), a state trajectory (dashed
red line) of the polarization vector $\pmb\mu+\mathrm{d\pmb\mu}$
(gray vector) is also shown due to a deviation $\mathrm{d}\omega_{x}$
(violet vector) on the component $\omega_{x}$ of the precession frequency
leading to $\omega_{x}+\mathrm{d}\omega_{x}$ (purple-red vector).
In (b) the polarization vector follows a trajectory driven by a coherent
evolution until a non-demolition measurement projects its component
on the z-axis. (c) Qubit-probe polarization $\mu_{z}$ along the $z$-axis
as a function of time $t$, for the coherent and projected evolutions
(red and blue color lines respectively) with $\omega_{x}$ and $\omega_{x}+\mathrm{d}\omega_{x}$
(solid and dashed lines respectively). We consider an offset $\omega_{z}$
that shows that the exponential decay resulting from a projected evolution
induces larger differences between the evolutions corresponding to
$\omega_{x}$ and $\omega_{x}+\mathrm{\mathrm{d}}\omega_{x}$, compared
to the differences induced by a coherent evolution. For the schematic
plots we use $\pmb\omega=2\pi(\cos\theta,0,\sin\theta)$ with $\theta=0.9\pi/2$
being $\omega_{z}\apprge6\omega_{x}$, $\mathrm{d}\omega_{x}=0.1\omega_{x}$,
$\tau=0.5$ and $\pmb\mu(0)=\mathbf{z}$.}
\label{fig:both_depic}
\end{figure}

The minimum attainable error of an unbiased estimation is determined
by the Quantum Cramer-Rao Bound \citep{Caves_1994_fisher,Paris2009_QUANTUM-ESTIMATION,escher2011general,amari_2016}
\begin{align}
\delta\omega_{x}^{2}=\expval{(\hat{\omega}_{x}-\omega_{x})^{2}}\geq\frac{1}{N\mathcal{F_{Q}}(\omega_{x})} & ,
\end{align}
where $\hat{\omega}_{x}$ is the estimated value, $N$ is the number
of performed measurements, and $\mathcal{F_{Q}}(\omega_{x})$ is the
Quantum Fisher Information (QFI) about $\omega_{x}$. The QFI can
be defined as
\begin{multline}
\mathcal{F_{Q}}(\omega_{x})=\sum_{n}\frac{(\partial_{\omega_{x}}\lambda_{n})^{2}}{\lambda_{n}}+\\
2\sum_{n\neq m}\frac{(\lambda_{n}-\lambda_{m})^{2}}{\lambda_{n}+\lambda_{m}}\left|\left\langle \lambda_{m}\right|\partial_{\omega_{x}}\left|\lambda_{n}\right\rangle \right|^{2},\label{eq:QFI_lambda-1}
\end{multline}
in the basis of the eigenvectors $\ket{\lambda_{n}}$ of the density
matrix $\rho$ and considering the corresponding eigenvalues $\lambda_{n}$
\citep{Paris2009_QUANTUM-ESTIMATION}. The QFI can be interpreted
as a metric that relates the square of an infinitesimal displacement
of the parameter $\mathrm{d}\omega_{x}^{2}$ and the square of a statistical
distance $\mathrm{d}s^{2}$ of the induced displacement on the quantum
state \citep{Caves_1994_fisher}
\begin{align}
\mathrm{d}s^{2}=\mathcal{F_{Q}}(\omega_{x})\mathrm{d}\omega_{x}^{2} & .\label{eq:displacement_metric}
\end{align}
In the case of a evolved state of the two level system driven by the
Hamiltonian of Eq. (\ref{eq:Hamiltonian}), an infinitesimal displacement
in the frequency $d\pmb\omega$ and polarization vector $d\pmb\mu$
is schematically shown in Fig. \ref{fig:both_depic}a.
 The QFI of Eq. (\ref{eq:QFI_lambda-1}) can be expressed in terms
of the polarization vector as
\begin{align}
\mathcal{F_{Q}}(\omega_{x})=\frac{1}{1-\mu^{2}}(\partial_{\omega_{x}}\pmb\mu_{\mathrm{r}})^{2}+(\partial_{\omega_{x}}\pmb\mu_{\mathrm{t}})^{2} & ,\label{eq:QFIfrompolars}
\end{align}
where $\partial_{\omega_{x}}\pmb\mu_{\mathrm{r}}$ and $\partial_{\omega_{x}}\pmb\mu_{\mathrm{t}}$
denote the radial and tangential components of the displacement $\partial_{\omega_{x}}\pmb\mu$
(see App. \ref{sec:QFI_polarization}). This means that information
is contained on changes on the absolute value of the polarization
vector and on its direction.

\section{\label{sec:QFI-coherent}Quantum Fisher Information for a qubit-probe
under coherent evolution}

We analyze here the estimation protocol when the qubit sensor undergoes
a coherent dynamic \citep{zwick16,zwick16crit,Degen2017,Poggiali2018,Zwick2019_mukherjee_enhanced,zwick2020precision,wang2020coherent,mu2020coherent,zwick2023quantum}.
The QFI of the coupling strength $\omega_{x}$, obtained by measuring
$\left\langle \sigma_{z}\right\rangle $ at time $t$ under a coherent
free evolution, is determined by the difference between the polarization
trajectories squematized in Figure \ref{fig:both_depic}, following
Eq. (\ref{eq:QFIfrompolars}). The dynamics of these polarization
trajectories of the qubit-probe state $\pmb\mu$ and $\pmb\mu+d\pmb\mu$,
correspond to precessions with frequencies $\pmb\omega$ and $\pmb\omega+d\pmb\omega$,
with the couplings $\omega_{x}$ and $\omega_{x}+\mathrm{d}\omega_{x}$
respectively (Fig. \ref{fig:both_depic}a). Figure \ref{fig:both_depic}c
shows the corresponding qubit-probe polarization $\mu_{z}$ as a function
of the evolution time. Since the polarization norm is conserved during
the coherent evolution, the first term in Eq. (\ref{eq:QFIfrompolars})
vanishes, and the QFI coincides with the squared euclidean displacement
induced on the polarization vector, i.e. the derivative $\partial_{\omega_{x}}\pmb\mu_{\mathrm{r}}(t)=0$
and thus the $\mathcal{\mathcal{F_{Q}^{\mathrm{coh}}}}(\omega_{x})=(\partial_{\omega_{x}}\pmb\mu_{\mathrm{t}}(t))^{2}$.
This displacement is induced by two contributions: one due to the
change in the norm of the precession frequency and the other due to
the change in the precession cone (the precession axis), as shown
in Fig. \ref{fig:both_depic}a. The later induces a periodic displacement,
becoming maximal at half periods of the precession and null at every
period, while the displacement induced by the former grows quadratically
as a function of time and defines the more significant contribution
to the QFI at long evolution times. More explicitly, the polarization
vector at the evolution time $t$ is
\begin{align}
\pmb\mu(t)=\mu_{0}\begin{pmatrix}\dfrac{\omega_{x}\omega_{z}}{\omega^{2}}(1-\cos(\omega t))\\
-\dfrac{\omega_{x}}{\omega}\sin(\omega t)\\
\dfrac{\omega_{z}^{2}}{\omega^{2}}+\dfrac{\omega_{x}^{2}}{\omega^{2}}\cos(\omega t)
\end{pmatrix} & ,\label{eq:muvst}
\end{align}
where $\omega=\abs{\pmb\omega}$. The spin-state polarization on the
longitudinal axis is
\begin{equation}
\mu_{z}(t)=\mu_{0}\alpha(t),\:\alpha(t)=\dfrac{\omega_{z}^{2}}{\omega^{2}}+\dfrac{\omega_{x}^{2}}{\omega^{2}}\cos(\omega t),\label{eq:muz}
\end{equation}
where $\mu_{0}$ is a constant giving the initial polarization. The
longitudinal polarization $\mu_{z}(t)$ oscillates with an amplitude
$\mu_{0}\omega_{x}^{2}/\omega^{2}$ with respect to its mean value
$\mu_{0}\omega_{z}^{2}/\omega^{2}$ as displayed in Fig. \ref{fig:both_depic}.

\begin{figure}
\begin{centering}
\includegraphics[width=0.85\columnwidth]{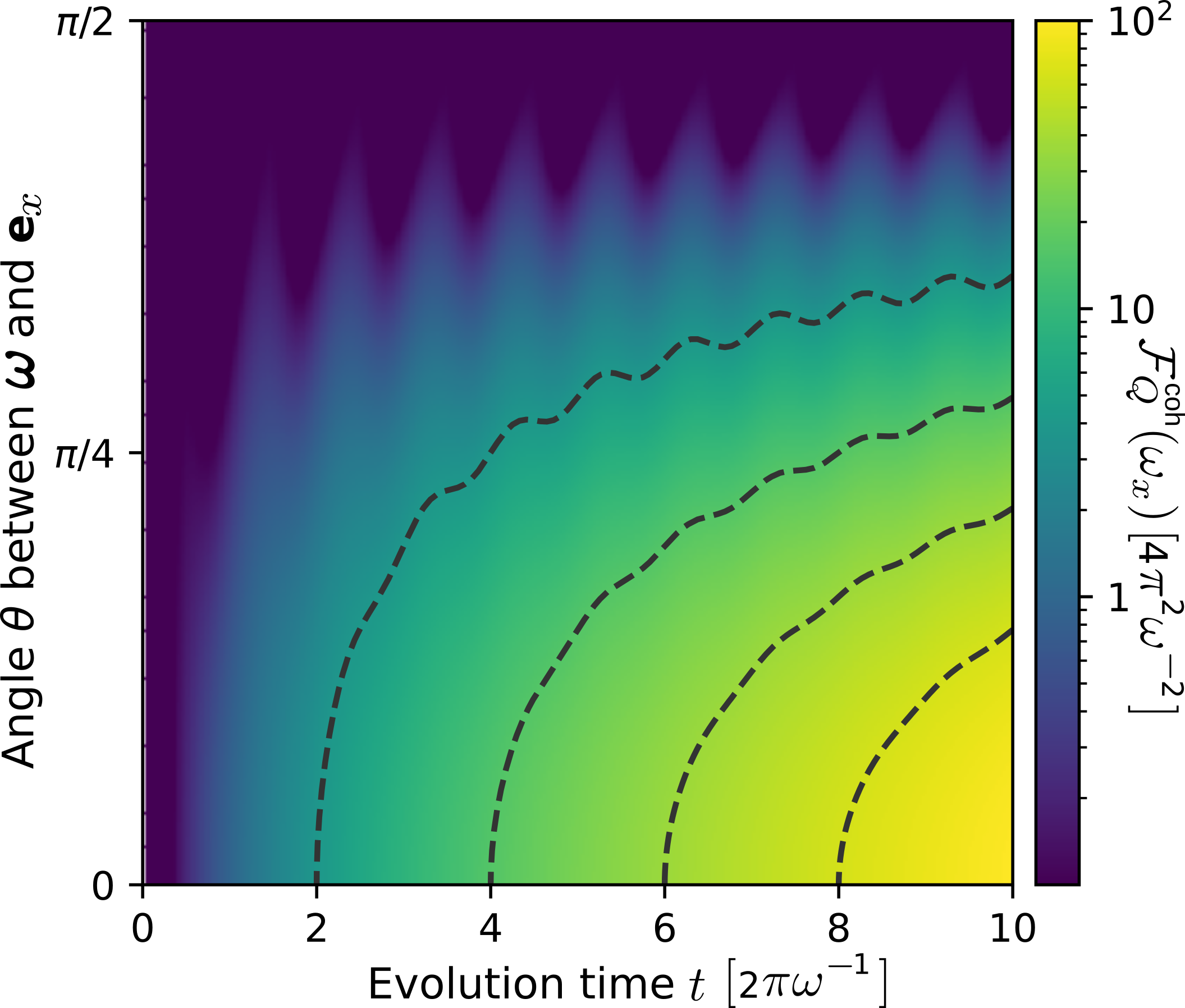}
\par\end{centering}
\caption{\label{fig:QFI vs theta vs t}Quantum Fisher Information of $\omega_{x}$
as a function of the evolution time $t$ and of the angle $\theta=\arctan\left(\frac{\omega_{z}}{\omega_{x}}\right)$
between the precesion frequency vector $\pmb\omega$ and the direction
$\mathbf{x}$ for a coherent evolution from the initial state $\pmb\mu(0)=\mathbf{z}$.
The dashed lines indicate contour lines that approximatelly satisfy
the relation $t\propto1/\cos^{2}\theta$.}
\end{figure}

Then, the Fisher information is

\begin{multline}
\mathcal{F_{Q}^{\mathrm{coh}}}(\omega_{x})=\mu_{0}^{2}\left(\dfrac{\omega_{x}^{2}}{\omega^{2}}t\begin{pmatrix}\dfrac{\omega_{z}}{\omega}\sin(\omega t)\\
-\cos(\omega t)\\
-\dfrac{\omega_{x}}{\omega}\sin(\omega t)
\end{pmatrix}\right.\\
\left.+\dfrac{\omega_{z}}{\omega^{2}}\begin{pmatrix}\dfrac{\omega_{z}^{2}-\omega_{x}^{2}}{\omega^{2}}(1-\cos(\omega t))\\
-\dfrac{\omega_{z}}{\omega}\sin(\omega t)\\
-2\dfrac{\omega_{x}\omega_{z}}{\omega^{2}}(1-\cos(\omega t))
\end{pmatrix}\right)^{2}\\
=\mu_{0}^{2}\dfrac{\omega_{x}^{4}}{\omega^{4}}t^{2}\left(1+\dfrac{2\omega_{z}^{2}\sin(\omega t)}{\omega_{x}^{2}\omega t}\right.\\
\left.+\left(\dfrac{\omega_{z}}{\omega_{x}^{2}t}\right)^{2}\left[(1-\cos(\omega t))^{2}+\dfrac{\omega_{z}^{2}}{\omega^{2}}\sin^{2}(\omega t)\right]\right),
\end{multline}
which for long times $t\gg\frac{\omega_{z}}{\omega_{x}^{2}}$, takes
the approximated form
\begin{equation}
\mathcal{F_{Q}^{\mathrm{coh}}}(\omega_{x})\approx\mu_{0}^{2}\left(\dfrac{\omega_{x}}{\omega}\right)^{4}t^{2},\label{eq:QFI_coh_approx}
\end{equation}
that grows quadratically with time. The accompanying factor $\left(\dfrac{\omega_{x}}{\omega}\right)^{4}$
is the fourth power of the cosine of the angle $\theta$ between the
precession frequency $\pmb\omega$ and the $\mathbf{x}$ direction,
$\omega_{x}=\omega\cos\theta$. The QFI as a function of $\theta$
and the evolution time shows approximatelly countourn lines given
by the relation $t\propto1/\cos^{2}\theta=1+\frac{\omega_{z}^{2}}{\omega_{x}^{2}}$,
as shown in Fig. \ref{fig:QFI vs theta vs t}. Thus, the QFI of $\omega_{x}$
grows faster as the on resonance condition is approached, and tends
to $0$ as $\theta\rightarrow\pi/2$ by increasing the offset.

\section{\label{sec:Quantum-Fisher-Information_projected-evolution}Quantum
Fisher Information for a qubit-probe under projected evolution}

We here analyze the estimation protocol when the qubit sensor undergoes
an incoherent dynamic via projected evolutions along the $\mathbf{z}$
axis. This is achieved through periodic projective –non-demolition–
measurements of the observable $\sigma_{z}$ at stroboscopic times
with a delay $\tau$ \citep{kwiat1999high,kofman00,alvarez_controlling_2012,zwick16,muller2016stochastic,muller2020noise,zwick2023quantum}.
Experimental realization of such measurements often entails applying
repetitive random magnetic field gradients \citep{Alvarez2010a,alvarez_controlling_2012}
or employing stochastic processes \citep{Dalibard1992,Nielsen1998,Teklemariam2003,Tycko2007,Zheng2013}
on the qubit-probe system. In the case of magnetic resonance, direct
projective measurements are challenging, but they can be replicated
using induced dephasing techniques \citep{Nielsen1998,Xiao2006,Alvarez2010a,Tycko2007,alvarez_controlling_2012,Zheng2013}.

A schematic representation of the probe state evolution is illustrated
in Fig. \ref{fig:both_depic}b. Immediately after the projective measurement,
the qubit-probe polarization lies on the $\mathbf{z}$ axis. The transversal
component is erased by the non-demolition measurement process, and
thus following Eq. (\ref{eq:muvst}) the polarization norm is reduced
by a factor $\alpha(\tau)$ as defined in Eq. (\ref{eq:muz}). For
an evolution time $t$ that contains $n$ stroboscopic projective
measurements, $t=n\tau+\Delta t$ with $n\in\mathbb{N}$ and $\Delta t\in[0,\tau)$,
the polarization $\mu_{z}$ is
\begin{equation}
\mu_{z}(t)=\mu_{0}\alpha(\tau)^{n}\alpha(\Delta t).\label{eq:muz_proj}
\end{equation}
 At the stroboscopic times $n\tau$, the polarization decays exponentially
with the characteristic time
\begin{align}
t_{c} & =-\frac{\tau}{\ln\abs{\alpha(\tau)}}.\label{eq:t_characteristic_stroboscopic}
\end{align}
The evolution of the polarization $\mu_{z}(t)$ is shown in Fig. \ref{fig:both_depic}c
for the precession frequencies $\pmb\omega$ and $\pmb\omega+d\pmb\omega$,
with the couplings $\omega_{x}$ and $\omega_{x}+\mathrm{d}\omega_{x}$
respectively as in the coherent evolution case. Notice that the projected
evolution difference between the two cases is significantly larger
than that for the coherent evolution case.
\begin{figure}
\centering

\includegraphics[width=0.9\columnwidth]{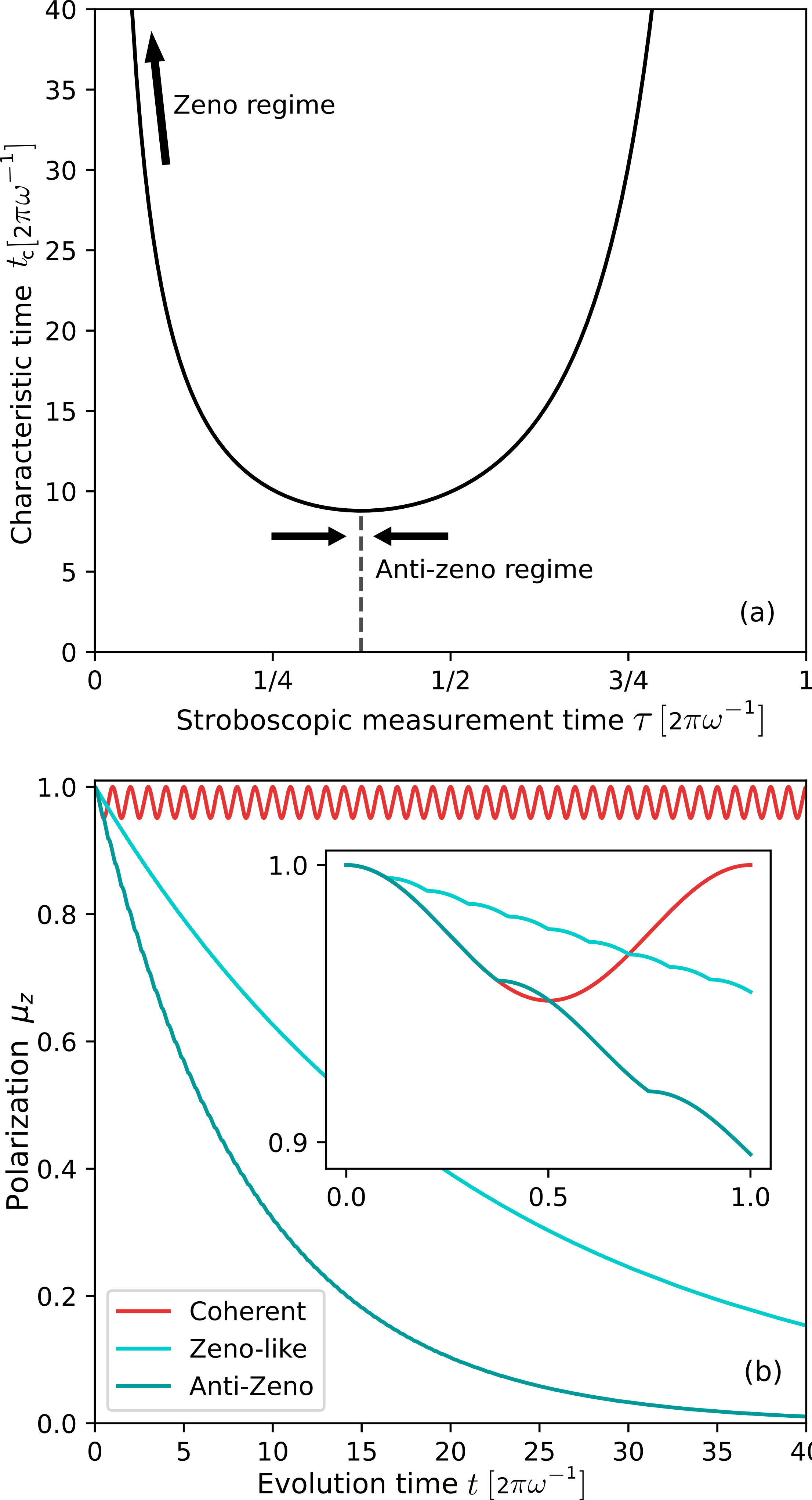}

\caption{(a) Characteristic time $t_{c}$ of a measurement-induced decay as
a function of the stroboscopic time $\tau$. The Zeno regime ocurrs
when $\tau$ tends to $0$, while the Anti-Zeno regime ocurrs for
the region where the characteristic time $t_{c}$ is minimum around
$\tau\approx\frac{3}{8}$. For $\omega_{z}\gg\omega_{x}$, $t_{c}\approx\frac{\omega^{2}\tau}{2\omega_{x}^{2}sin^{2}(\frac{\omega\tau}{2})}$,
and the minimum ocurr for $\omega\tau=\text{\ensuremath{\tan}}(\frac{\omega\tau}{2})$.
(b) Polarization $\mu_{z}$ along the $z$-axis as a function of the
total evolution time, for coherent and two projective evolutions corresponding
to a Zeno ($\tau=0.1$) and Anti-Zeno regime ($\tau\approx\frac{3}{8}$)
(red, blue and cyan color lines respectivesly). Here $\pmb\omega=2\pi(\cos\theta,0,\sin\theta)$
with $\theta=0.9\pi/2$ being $\omega_{z}\apprge6\omega_{x}$, $\pmb\mu(0)=\mathbf{z}$.}
\label{fig:charactime_vs_tau}
\end{figure}

Figure \ref{fig:charactime_vs_tau}a shows the decay time $t_{c}$
as a function of the stroboscopic measurement time $\tau$ for $\omega_{z}\apprge6\omega_{x}$.
A similar qualitative behavior is seen in general for $\omega_{z}>\omega_{x}$,
as the polarization state $\pmb\mu$ does not cross the $\mu_{z}=0$
plane during its evolution. As the stroboscopic time $\tau\rightarrow0$,
the evolution of the polarization factor $\alpha(\tau)\rightarrow1$
and thus the characteristic decay time $t_{c}\rightarrow\infty$
defining the Zeno regime where the QZE is manifested \citep{misra77,kofman00}.
Figure \ref{fig:charactime_vs_tau}a also shows a region for values
of $\tau$, where the decay time $t_{c}$ is minimal corresponding
to a Anti-Zeno regime \citep{kofman00}. These Zeno and Anti-Zeno
regimes induce the slowest and fastest exponential decay of the polarization
$\mu_{z}$ evolution under stroboscopic measurements as shown in Fig.
\ref{fig:charactime_vs_tau}b.

We now calculate the QFI from Eq. (\ref{eq:QFIfrompolars}). Due to
the reduction of the polarization norm by the factor $\alpha(\tau)$
at every projective measurement, the first term in Eq. (\ref{eq:QFIfrompolars})
becomes a piecewise constant evolution. The coherent evolution between
the projective measurements contributes to the QFI via the second
term of Eq. (\ref{eq:QFIfrompolars}) with a value equal to the one
of a coherent evolution after the projection. Thus the second term
of the QFI is a coherent contribution that only adds a small quantity
on top of the main overall information given by the first term due
to the incoherent evolution counterpart. If the stroboscopic time
$\tau$ is small, $\omega\tau\ll2\pi$, the first term of the QFI
is generated by a decaying incoherent evolution of the qubit-state
$\mu_{z}$. The QFI at the stroboscopic times $t=n\tau$ is given
only by the radial –first– term in Eq. (\ref{eq:QFIfrompolars}).
This is regardless of the value of $\omega_{x}$, as all trajectories
at those specific times are along the $\mathbf{z}$ axis (see Fig.
\ref{fig:both_depic}b). Furthermore, during the coherent evolution
between the projective measurements, since the polarization vector
conserves its norm, the radial term remains constant, and the second
term takes the usual form for a coherent evolution as described in
Sec. \ref{sec:QFI-coherent}, with an initial polarization given by
the one obtained at the last projective measurement. Hence, the QFI
may be approximated by the first term of Eq. (\ref{eq:QFIfrompolars})
up to a difference of $\mu_{0}^{2}\alpha(\tau)^{2n}\Delta t^{2}$,
where $\Delta t$ is the elapsed time after the last measurement.
Considering this approximation, the QFI takes the form (see App.
\ref{ap1qfi})
\begin{align}
\mathcal{F_{Q}^{\mathrm{proj}}}(\omega_{x}) & =t^{2}\mu_{0}^{2}\frac{1}{\tau^{2}}\frac{\alpha(\tau)^{2(t/\tau-1)}}{1-\mu_{0}^{2}\alpha(\tau)^{2t/\tau}}(\partial_{\omega_{x}}\alpha(\tau))^{2}.\label{eq:QFI-proj}
\end{align}

\begin{figure}
\centering \includegraphics[width=0.93\columnwidth]{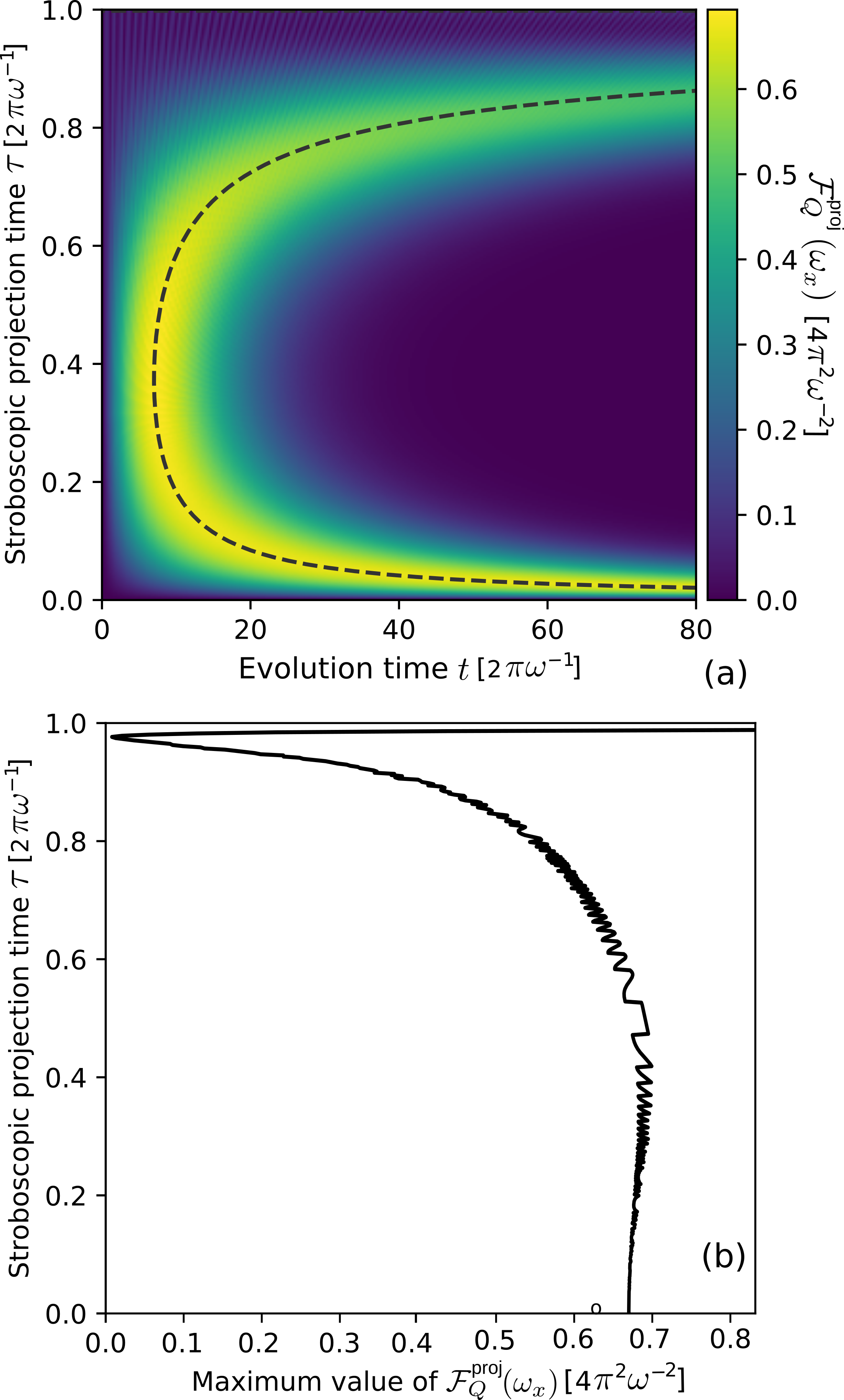}\caption{Quantum Fisher information of the coupling strength $\omega_{x}$
for the projected evolution of the quantum-probe. (a) The QFI $\mathcal{F_{Q}}(\omega_{x})$
as a function of the stroboscopic measurement time $\tau$ and the
total evolution time $t$. The dashed line shows the maximum value
of the QFI $\mathcal{F_{Q\mathrm{,max}}^{\mathrm{proj}}}=\mathcal{\mathcal{F_{Q}^{\mathrm{proj}}}}(t_{\text{{max} }})$
obtained at $t_{\text{{max}}}$. (b) Maximum QFI $\mathcal{F_{Q\mathrm{,max}}^{\mathrm{proj}}}$
as a function of $\tau$. The vertical axes in both panels represents
the stroboscopic time $\tau$. Parameters used for the plot: $\pmb\omega=2\pi(\cos\theta,0,\sin\theta)$
with $\theta=0.9\pi/2$ being $\omega_{z}\apprge6\omega_{x}$ and
$\pmb\mu(0)=\mathbf{z}$. The oscillations in (b) are due to the finite
numerical calculations.}
\label{fig:QFI_vs_theta_vs_t_proj}
\end{figure}

The QFI as a function of time for this projected evolution is represented
by a family of self-similar functions parameterized by $\pmb\omega$
and $\tau$. Figure \ref{fig:QFI_vs_theta_vs_t_proj} shows this QFI
of the coupling strength $\omega_{x}$ as a function of stroboscopic
time $\tau$ and the total evolution time $t$. The dashed line in
Fig. \ref{fig:QFI_vs_theta_vs_t_proj}(a) shows the maximum value
of the QFI $\mathcal{F_{Q\mathrm{,max}}^{\mathrm{proj}}}=\mathcal{\mathcal{F_{Q}^{\mathrm{proj}}}}(t_{\text{{max} }})$
obtained at $t_{\text{{max} }}$ that satisfies $\partial_{t}\mathcal{\mathcal{F_{Q}^{\mathrm{proj}}}}(t_{\text{{max} }})=0$.
The maximum QFI ocurr at
\begin{align}
t_{\text{{max} }} & =\xi(\mu_{0})t_{c}, & \xi(\mu_{0}) & =1+\frac{1}{2}\operatorname{W}(-2\mu_{0}^{2}/\mathrm{e^{2}}),\label{eq:tmax-1}
\end{align}
where $\operatorname{W}$ is the principal branch of the Lambert's
W function and vary monotonically from $0.79$ to $1$ as $\mu_{0}$
goes from $1$ to $0$. Then the maximum QFI is given by
\begin{align}
\mathcal{\mathcal{F_{Q\mathrm{,max}}^{\mathrm{proj}}}} & =\varphi(\mu_{0})\frac{t_{c}^{2}}{\alpha(\tau)^{2}}\frac{\omega_{x}^{2}}{\omega^{2}}\left[2\frac{1-\cos(\omega\tau)}{\omega\tau}\frac{\omega_{z}^{2}}{\omega^{2}}+\sin(\omega\tau)\frac{\omega_{x}^{2}}{\omega^{2}}\right],\label{eq:Hmax-1}\\
\varphi(\mu_{0}) & =\mu_{0}^{2}\frac{(\xi(\mu_{0}))^{2}}{\mathrm{e^{2\xi(\mu_{0})}-\mu_{0}^{2}}},\nonumber 
\end{align}
and the behaviour of $F_{Q\mathrm{,max}}^{\mathrm{proj}}(\omega_{x})$
as a function of $\tau$ is shown in Fig. \ref{fig:QFI_vs_theta_vs_t_proj}(b).

Approaching the Zeno regime $\omega\tau\rightarrow0$, $\alpha(t)\approx1-\omega_{x}^{2}t^{2}$
and the QFI $\mathcal{F_{Q\mathrm{,max}}^{\mathrm{proj}}}\rightarrow4\varphi(\mu_{0})/\omega_{x}^{2}$
is a constant value that is obtained at $t_{\text{{max} }}\approx\frac{\xi(\mu_{0})}{\omega_{x}^{2}\tau}$,
where both are independent of $\omega_{z}$. This is an important
feature of the QZE estimation strategy, as it does not require previous
knowledge of the offset $\omega_{z}$ to make an efficient inference.
For large offsets $\omega_{z}\gg\omega_{x}$, this constant value
for $\mathcal{F_{Q\mathrm{,max}}^{\mathrm{proj}}}$ is extended to
the Anti-Zeno regime requiring less total evolution time to be achieved
as can be observed in Fig. \ref{fig:QFI_vs_theta_vs_t_proj}. Notice
that outside the QZE regime, previous knowledge of the offset is needed
for an efficient estimation of the coupling. By continuing to increase
$\tau$, the $\mathcal{F_{Q\mathrm{,max}}^{\mathrm{proj}}}$ decreases
to zero. This could be related to the manifestation of critical phenomena
in the extractable information defining transitions between dynamic
regimes of the sensor \citep{zwick16crit}. Then, $\mathcal{F_{Q\mathrm{,max}}^{\mathrm{proj}}}$
rapidily increases as the evolution approaches the coherent regime,
i.e. for $\tau\rightarrow2\pi\omega^{-1}$, but it requires a total
evolution time orders of magnitude larger than its Anti-Zeno counterpart.

It is worth noting that the functional form of $\mathcal{F_{Q\mathrm{,max}}^{\mathrm{proj}}}$
differs from that of the level curves for the QFI in a coherent evolution,
as illustrated in Fig. \ref{fig:QFI vs theta vs t} compared to Fig.
\ref{fig:QFI_vs_theta_vs_t_proj}. Specifically, for sufficient large
values of $\omega_{z}$, $\mathcal{F_{Q\mathrm{,max}}^{\mathrm{proj}}}$
becomes constant and defines a contour line that gives the higher
information gain at an earlier time compared to the corresponding
contour line determined from a coherent evolution.

\section{\label{sec:Maximizing-Information-with}Maximizing Information with
the Quantum-Zeno Effect}

\begin{figure}
\centering \includegraphics[width=0.9\columnwidth]{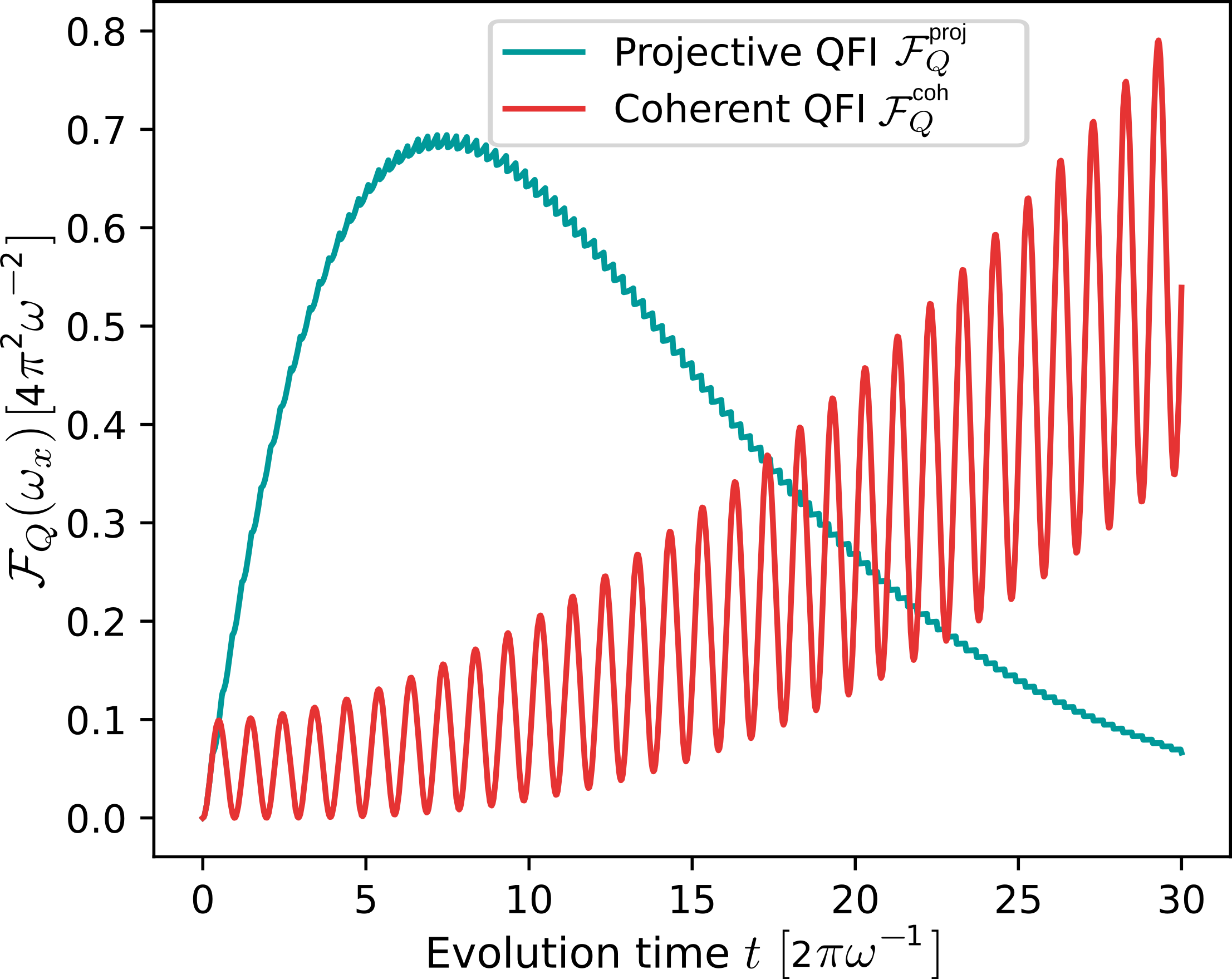}

\caption{QFI for a large offset $\omega_{z}\apprge6\omega_{x}$ for both coherent
and projective evolution estimation. Although the QFI extracted from
a coherent evolution yield higher information at long time, the QFI
extracted from projective evolution yields a higher value at shorter
times. The parameter used for the plot are: $\pmb\omega=2\pi(\cos\theta,0,\sin\theta)$
with $\theta=0.9\pi/2$ being $\omega_{z}\apprge6\omega_{x}$, $\tau=0.3$
and $\pmb\mu(0)=\mathbf{z}$.}
\label{fig:proj_winning_QFI}
\end{figure}

Here, we compare the estimation efficiency of the coupling strength
$\omega_{x}$, when the qubit-probe undergoes coherent and projected
(incoherent) evolutions. In the off-resonance regime ($\omega_{z}>\omega_{x}$,)
the difference between the evolution trajectories of the observable
$\mu_{z}$ for a small deviation on the precession frequency $\omega_{x}$
becomes barely distinguishable during coherent evolution (see Fig.
\ref{fig:both_depic}c). On the contrary, projected evolutions leads
to larger differences between these trajectories in a region of times
before the decaying signal is lost (see Fig. \ref{fig:both_depic}c).
As previously discussed in Eq. (\ref{eq:displacement_metric}), the
differences observed in the trajectories resulting from a small deviation
in the parameter $\omega_{x}$ can provide valuable information about
it. Figure \ref{fig:proj_winning_QFI} compares the time-dependent
QFI of $\omega_{x}$ between a coherent ($\mathcal{F_{Q}^{\mathrm{coh}}})$
and projected ($\mathcal{F_{Q}^{\mathrm{proj}}}$) evolution estimation
process, illustrating a representative functional behavior for a case
with a large offset. 
It reflects the key predicted results of this article: the QFI $\mathcal{F_{Q}^{\mathrm{proj}}}$
extracted from projected evolutions achieves higher values at shorter
times compared to the the QFI information $\mathcal{F_{Q}^{\mathrm{coh}}}$
extracted from a coherent free evolution when the offset is large
$\omega_{z}>\omega_{x}$.

The $\mathcal{F_{Q}^{\mathrm{proj}}}$reaches a maximum and then it
decays, while $\mathcal{F_{Q}^{\mathrm{coh}}}$ continuously grows
as a function of time, as predicted in Eq. (\ref{eq:QFI_coh_approx}).
This happens because the observable longitudinal polarization $\mu_{z}$
exponentially decays to zero under projective evolutions, thus reducing
the QFI after some time, while the coherent evolution preserve the
polarization magnitude along time thus allowing to increse the QFI
indefinitely. However, the QFI can indefinitely grow only under very
ideal conditions, where the qubit-probe does not suffer decoherence.

In a realistic experiments, there is always decoherent effects that,
depending the setup, can be characterized by the relaxation times
$T_{1}$ or $T_{2}$, which are typically comparable to the precession
period $2\pi\omega^{-1}$ or one/two orders of magnitude larger. This
relaxation reduces the polarization $\mu_{z}$ by an exponential decaying
factor $e^{-t/T_{1}}$ or $e^{-t/T_{2}}$ respectively. Therefore
in general the maximal QFI that can be achieved for given values of
$\omega_{x}$ and the offset $\omega_{z}$, is obtained by an optimal
tradeoff between the ideal information gain from the qubit-probe evolution
and the total available time determined by sources of relaxation.
Thus, due to decoherence effects on the qubit-probe, achieving the
maximal QFI in the shortest possible time is crucial.

We thus determine the maximum attainable QFI within the time interval
$[0,\,t]$ for the coherent $F_{Q}^{\mathrm{coh}}$ and the projected
$F_{Q}^{\mathrm{proj}}$ evolution estimations. We compare them with
the quotient

\begin{align}
\frac{\mathcal{\mathcal{F_{Q\mathrm{,max}}^{\mathrm{proj}}}}}{\mathcal{F_{Q\mathrm{,max}}^{\mathrm{coh}}}} & =\frac{4\varphi(\mu_{0})}{\mu_{0}^{2}}\frac{\omega^{4}}{\omega_{x}^{6}}\frac{1}{t^{2}},\label{eq:quotient}
\end{align}
considering the approximations derived above, where $\mathcal{F_{Q\mathrm{,max}}^{\mathrm{coh}}}(\omega_{x})\approx\mu_{0}^{2}\left(\dfrac{\omega_{x}}{\omega}\right)^{4}t^{2}$
for $t\gg\frac{\omega_{z}}{\omega_{x}^{2}}$ and $\mathcal{F_{Q\mathrm{,max}}^{\mathrm{proj}}}\approx4\varphi(\mu_{0})/\omega_{x}^{2}$
for large offsets $\omega_{z}\gg\omega_{x}$ and $\tau\lesssim0.6$.
This expression thus defines the conditions for the optimal estimation
at the total available evolution time $t$. The estimation procedure
based on projective measurements on the qubit-probe maximize the information
about $\omega_{x}$ when the available time is

\begin{align}
t & <\frac{2\sqrt{\varphi(\mu_{0})}}{\mu_{0}\omega_{x}}\left(1+\frac{\omega_{z}^{2}}{\omega_{x}^{2}}\right).\label{eq:condition-largewz}
\end{align}
In the limit of low polarization $\mu_{0}\ll1$, $\varphi(\mu_{0})=e^{-2}\mu_{0}^{2}$,
and Eq. (\ref{eq:condition-largewz}) aproximates to 
\begin{equation}
t<\frac{2e}{\omega_{x}}\left(1+\frac{\omega_{z}^{2}}{\omega_{x}^{2}}\right),\label{eq:condition-low-polarization}
\end{equation}
which is independent of the particular initial polarization $\mu_{0}$.

\begin{figure}
\centering \includegraphics[width=0.98\columnwidth]{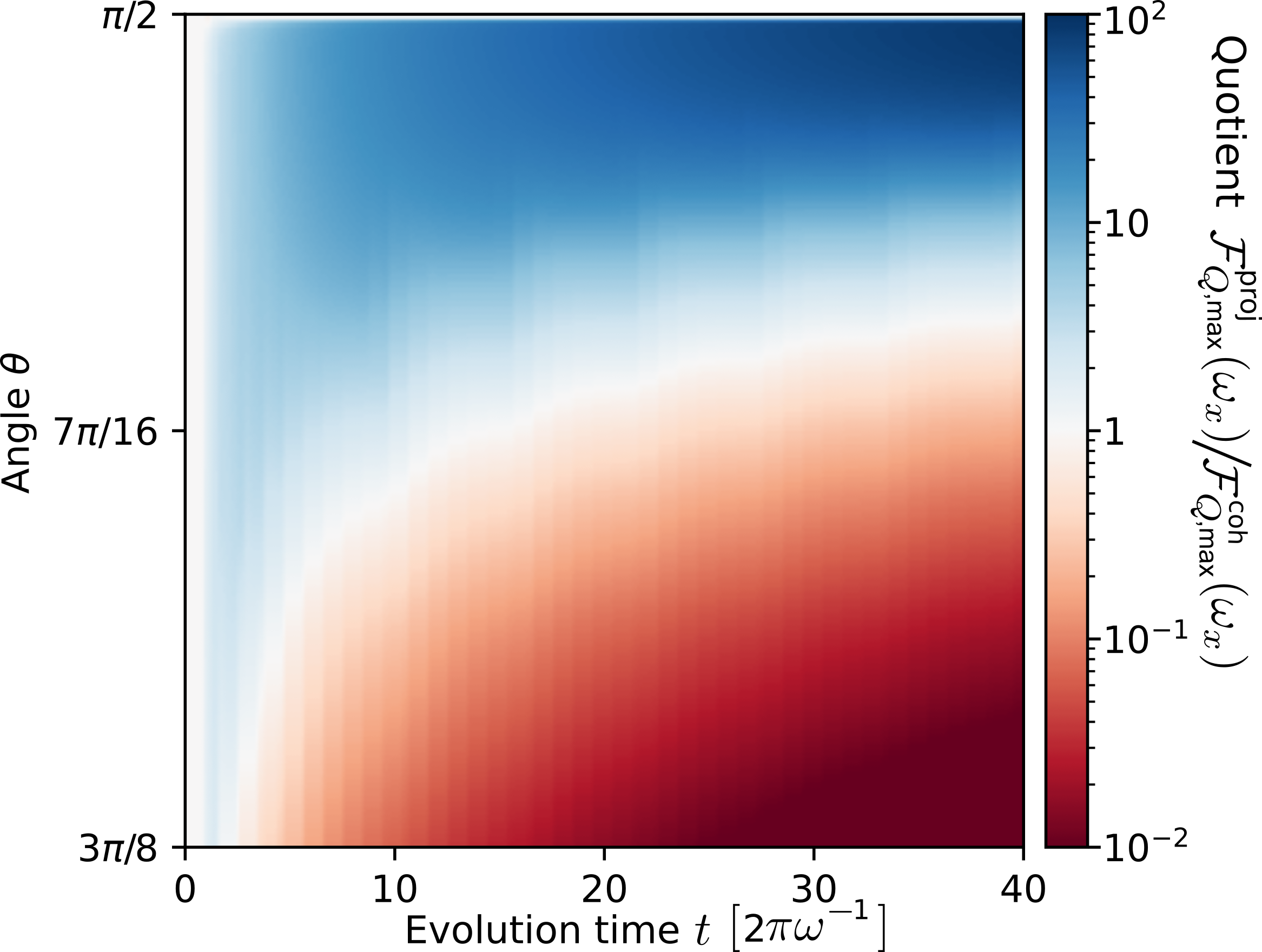} \caption{Quotient of the maximum obtainable QFI for projective and coherent
estimation as a function of the angle $\theta=\arctan\left(\frac{\omega_{z}}{\omega_{x}}\right)$
of $\pmb\omega$ with respect to $\mathbf{z}$ and total available
time $t$. The white color region corresponds to equal QFI that divides
the region where either estimation protocol is better.}
\label{fig:QFI_vs_theta_vs_t_comp}
\end{figure}

Figure \ref{fig:QFI_vs_theta_vs_t_comp} shows the quotient of Eq.
(\ref{eq:quotient}) as a function of the angle $\theta=\arctan\left(\frac{\omega_{z}}{\omega_{x}}\right)$
of $\pmb\omega$ with respect to the $x$ axis and the total available
time $t$. Areas colored in white indicate when the quotient is equal
to $1$, thereby defining the boundary between the parametric region
where projected or coherent evolution is more efficient por the parameter
estimation. For values of $\omega_{z}$ where the quotient is above
$1$, the projected evolution estimation is expected to perform better,
while the opposite is expected for values below it. This dividing
boundary curve is approximated by the relation $\omega_{z}\propto\omega_{x}\sqrt{\omega_{x}t-1}$,
provided that $t\geq t_{\text{{max}}}$. For values of $\omega_{z}$
above this boundary region, the offset is so strong that the contribution
of $\omega_{x}$ to the oscillatory dynamics requires more time than
is available. Consequently, coherent evolution does not provide a
better estimation compared to that extracted from the incoherent decaying
dynamics of the qubit-probe polarization. When the offset is below
this boundary region, the oscillatory coherent evolution of the quantum-probe
is sensitive enough to provide a larger QFI than the maximum attainable
by the incoherent evolution.

\section{\label{sec:Applications}Applications}

In this section we present physical examples that illustrate the advantages
of implementing the QZE inference protocol over the coherent strategy.
We begin by demonstrating the direct application of inferring AC magnetic
fields that are off resonant with the qubit-probe. Subsequently, we
explore how the same concept can be exploited for inferring spin-spin
couplings. In the latter case, we start with the simplest case of
a two-spin coupled system, followed by an example illustrating its
extension to a three spin system. Finally, we generalize the approach
to include many-spins.

The incoherent control necessary to achieve the Zeno effects may be
applied through quantum non-demolition (QND) projective measurements
\citep{zwick16}, as discussed in previous sections, and/or by induced
dephasing that mimics QND measurements \citep{Alvarez2010a,alvarez_controlling_2012,Nielsen1998,Cory2000,Xiao2006,Zheng2013}.
This induced dephasing, which repetitively and stroboscopically projects
the qubit state, can be achieved using various methods, including
random magnetic field gradients \citep{Alvarez2010a,alvarez_controlling_2012,Xiao2006},
$T_{2}$ relaxation \citep{Nielsen1998}, or stochastic interactions
\citep{Tycko2007,Dalibard1992,Teklemariam2003}. It is important to
note that the results of QND measurements do not necessitate a readout,
as these measurements solely guide the evolution of the qubit-probe
state.

\subsection{Sensing an off-resonant, external AC magnetic field with a 1/2-spin}

We consider a $1/2$-spin interacting with a homogeneous and static
magnetic field $\mathbf{B}_{0}=B_{0}\mathbf{z}$ as the qubit-probe.
It thus precesses with the Larmor frequency $\omega_{0}=\gamma B_{0}$
around the $\mathbf{z}$ axis, where $\gamma$ is the gyromagnetic
ratio. The qubit-probe is treated as a magnetometer for estimating
an AC magnetic field along the $\mathbf{x}$ axis, determined by $\mathbf{B}_{1}(t)=2B_{1}\cos(\omega t)\mathbf{x}$,
where $\omega$ is the carrier frequency and $B_{1}$ is the field
strength \citep{Degen2017,boss2017quantum,schmitt2017submillihertz,glenn2018high,jiang2023quantum,segawa2023nanoscale}.
In the rotating frame precessing at the angular frequency $\omega$,
and under the rotating wave approximation \citep{slichter}, the spin
interacts with an effective magnetic field $(\omega_{0}-\omega)/\gamma\mathbf{z}+B_{1}\mathbf{x}$.
Here, we have assumed that $B_{0}\gg B_{1}$ and that the offset $\omega_{z}=\omega_{0}-\omega$
is much lower than the Larmor frequency $\left|\omega_{z}\right|\ll\omega_{0}$.

This interaction can thus be map to the general two-level system Hamiltonian
of Eq. (\ref{eq:Hamiltonian}), where $\omega_{x}=\gamma B_{1}$,
$\omega_{y}=0$ and the offset $\omega_{z}=\gamma B_{0}-\omega$ defines
how far the AC field is from the on-resonance condition.

Following the results derived in Eqs. (\ref{eq:quotient})-(\ref{eq:condition-low-polarization}),
we can determine when the projective measurements protocol for estimating
the AC field strength $\omega_{x}=\gamma B_{1}$ provides a better
approach compared to a coherent evolution if the available total measurement
time is $t$. We thus obtain that if $t$ satisfies 
\begin{equation}
t<2e\frac{B_{1}^{2}+(B_{0}-\omega/\gamma)^{2}}{B_{1}^{3}},\label{eq:criteria_best estimation with QZE}
\end{equation}
projective measurement outperforms a coherent evolution estimation
protocol. This regime is useful when the on-resonance condition cannot
be fulfilled, which might occur when either $\omega$ or $B_{0}$
cannot be controlled to put the qubit-probe on resonance. For example,
this circumstance could arise when using an ensemble of spins that
feel different magnetic fields $B_{0}$, leading to an intrinsic large
offset value. This condition can arise in Nuclear Magnetic Resonance
(NMR) experiments using a mouse magnet that is placed at the side
or near the surface of the sample to be studied, causing the $B_{0}$
field to be spacially inhomogeneous within the sample, leading to
a broadening of the NMR spectrum \citep{juanperlo05}.

\subsection{Sensing spin-spin couplings}

The estimation of the coupling strength between interacting spins
is a crucial experimental challenge in chemistry for characterizing
molecular topology and various quantum technologies \citep{Braunschweiler1983,Caravatti1983,Ernst1990,Luy1999,Tycko2007,alvarez_controlling_2012,Shi2013,Degen2017}.
This is particularly relevant for Hamiltonian characterization \citep{Caravatti1983,Ernst1990,Tycko2007,segawa2023nanoscale,Degen2017,alvarez_controlling_2012}
as well as the improvement of spectroscopy techniques \citep{PhysRevX2015.5.041016,Smith2012,jiang2023quantum,schmitt2017submillihertz,glenn2018high,boss2017quantum},
optimization of hyperpolarization in NV centers \citep{king2015room,suter_rao2020level,wang2013sensitive,alvarez_local_2015,broadway2016anticrossing,ajoy2018orientation,Zangara2019}
and NMR \citep{pravdivtsev2013exploiting,pravdivtsev2014exploiting,ivanov2014role,pravdivtsev2014highly},
cross-polarization \citep{chattah03,alvarez_environmentally_2006,Hirschinger_crosspolarization_2011,Frydman_Efficient_2012,Hirschinger_crosspolarization_2013,raya2017sensitivity,hirschinger2023analytical},
the characteriation of molecular structures with NMR that define physicochemistry
properties or inter-nuclear distances \citep{slichter,ernst74,alvarez_controlling_2012,Shi2013},
among others.

\subsubsection{\label{subsec:Two-spin-coupling}Two-spin coupling}

To illustrate the introduced inference method, we focus on estimating
the coupling strength and inferring the Hamiltonian of two interacting
spins systems. Specifically, we consider a two-spin coupled system
during a cross-polarization experiment in NMR \citep{hartmannhahn60,ernst74}.
This technique is useful for transferring magnetization from a system
with high abundance and/or polarization (spin $I$), such as $^{1}H$,
to a system of low abundance and/or small gyromagnetic factor $\gamma$
(spin S), like $^{13}C$. The two spin species are subjected to a
static magnetic field $B_{0}$ as in the previous example.

The Zeeman interaction defines the resonance frequencies of each spin
$\omega_{0,\,i}=\gamma_{i}B_{0}$ where $i=I,S$, typically in the
order of hundreds of MHz and differing also in that order. The dipolar
interaction between the spins is in the order of kHz, making the polarization/magnetization
exchange between them negligible. To generate the polarization exchange
(cross polarization), it is necessary to put the two spin species
on resonance. For that, oscillating magnetic fields $\mathbf{B}_{1,i}(t)$
of frequences $\omega_{0,\,i}$ ($i=I,S$) are applied. In the high
radio frequency field regime where $\left|\omega_{1,\,I}+\omega_{1,\,S}\right|\gg\left|b\right|$
with $\omega_{1,\,i}=\gamma_{i}B_{1,i}$ and $B_{1,i}=\max\left|\mathbf{B}_{1,i}(t)\right|$,
a secular approximation can be done \citep{ernst74}. This yields
the interacting Hamiltonian in the double rotating frame precessing
at the two spin frequencies

\begin{align}
H & =-\frac{1}{2}\Delta\left(S_{z}-I_{z}\right)+b\left(S_{x}I_{x}+S_{y}I_{y}\right).\label{eq:Hamilt-2spin-interacting}
\end{align}
Here $I_{v}=\frac{1}{2}\sigma_{v,I}$ and $S_{v}=\frac{1}{2}\sigma_{v,S}$
are the spin operators with $v=x,y,z$, $\Delta=\left(\omega_{1,\,S}-\omega_{1,\,I}\right)$
is the off-resonant energy, and the dipolar interaction is 
\begin{align}
b & =-\frac{1}{2}\left(\frac{\mu_{0}\gamma_{I}\gamma_{S}}{4\pi r^{3}}\right)\frac{3r_{z}^{2}-r^{2}}{r^{2}},
\end{align}
where $r$ is the modulus of the internuclear distance vector, and
$r_{z}$ its $z$ component. Then, the on-resonance condition for
achieving a full cross-polarization, called the Hartmann-Hahn condition
\citep{hartmannhahn60}, is $\Delta=0$.

The sample is initially polarized by the action of the static field
$B_{0}\mathbf{z}$. The polarization in the insensitive species $S$
is typically negligible or removed for quantitative analisis, so only
the sensitive species $I$ is polarized along $\mathbf{z}$-axis according
to the Boltzmann distribution. The initial state is diagonal in the
Zeeman basis, and as the Hamiltonian of Eq. (\ref{eq:Hamilt-2spin-interacting})
conserves the total magnetization on the $\mathbf{z}$-axis, the evolution
of the spins takes place within density matrix blocks conserving the
total magnetization $S^{z}+I^{z}$. The matrix blocks of the spaces
$\{{\uparrow\uparrow}\}$ and $\{{\downarrow\downarrow}\}$ do not
generate dynamics in the system, so their populations are constant
over time. The dynamics ocurr only on the subspace of the Zeeman states
$\{\uparrow\downarrow,\,\downarrow\uparrow\}$, subject to the effective
Hamiltonian 
\begin{align}
\frac{1}{2}\begin{pmatrix}-\Delta & b\\
b & \Delta
\end{pmatrix}.
\end{align}
 This Hamiltonian is equivalent to Eq. (\ref{eq:Hamiltonian}), with
an angular frequency 
\begin{align}
\pmb\omega & =\frac{1}{\hbar}(-b,\,0,\,\Delta),
\end{align}
with a pauli operator $\pmb\sigma_{\uparrow\downarrow}$ acting on
the two state subspace $\{\ket{\uparrow\downarrow},\,\ket{\downarrow\uparrow}\}$.
Then, the process to estimate the coupling strength $b$ is analogous
to the estimation of $\omega_{x}$ with an offset $\omega_{z}=\Delta$
as discussed in the previous sections.

Since the blocks of the spaces $\{\uparrow\uparrow\}$ and $\{\downarrow\downarrow\}$
are static, they do not contribute to the QFI of any parameter. Only
the state in the block of space $\{\uparrow\downarrow,\,\downarrow\uparrow\}$
contributes to the QFI. In this two-level space, whith a trace of
$\frac{1}{2}$, the polarization vector is defined by 
\begin{align}
\rho_{\uparrow\downarrow} & =\frac{1}{4}\left[\mathrm{I}_{\uparrow\downarrow}+\pmb{\mathfrak{\mu}}\cdot\pmb\sigma_{\uparrow\downarrow}\right],
\end{align}
where the subscript $\uparrow\downarrow$ indicates that the operators
act on the subspace generated by $\{\ket{\uparrow\downarrow},\,\ket{\downarrow\uparrow}\}$.
The initial condition is
\begin{align}
\pmb{\mu}(t=0) & =\left(\frac{\exp\left(-\frac{1}{2}\frac{\omega_{0,\,I}}{kT}\right)}{\cosh\left(-\frac{1}{2}\frac{\omega_{0,\,I}}{kT}\right)}-1\right)\,\mathbf{z}\\
 & \approx-\frac{1}{2}\frac{\omega_{0,\,I}}{kT}\,\mathbf{z},
\end{align}
where this last approximation corresponds to the high temperature
limit, valid for NMR experiments at room temperature \citep{slichter}.
Thus the polarization vector precesses with the angular frequency,
leading to an evolution for its $z$ component given by
\begin{align}
\mathfrak{\mu}_{z}(t) & =\frac{1}{2}\mu_{0}\frac{\Delta^{2}+b^{2}\cos(\sqrt{b^{2}+\Delta^{2}}t)}{\Delta^{2}+b^{2}}.
\end{align}
This polarization component represents the magnetization exchange
between the spins, whose polarization transfer amplitude is $b^{2}/\left(\Delta^{2}+b^{2}\right)$.
Therefore, an off-resonant offset $\Delta$ far from the Hartmann-Hahn
condition reduces the information gain for the estimation of $b$
based on the free coherent evolution of the system. For an off-resonant
exchange of polarization, with large offset $\Delta$, projective
measurements applied on the spins on the $\text{z}$ axis lead to
a more efficient estimation of the coupling strength $b$ if the available
total evolution time satisfies
\begin{align}
t & <2\mathrm{e}\frac{b^{2}+\Delta^{2}}{b^{3}}=2\mathrm{e}\frac{b^{2}+\left(\omega_{1,\,I}+\omega_{1,\,S}\right)^{2}}{b^{3}},
\end{align}
according to Eqs. (\ref{eq:quotient})-(\ref{eq:condition-low-polarization}).
Typically the coherent evolution decoheres due to interactions with
the environment, imposing restrictions to the available time for the
inference \citep{chattah03,alvarez_environmentally_2006,Alvarez2007}.

The impossibility of generating the Hartmann-Hahn condition occurs
in many experimental situations, particularly in cases of very complex
spectra, such as spin ensembles with different resonance frequencies,
as is the case of solid-state systems with polycrystalline samples
\citep{Hirschinger_crosspolarization_2011,Frydman_Efficient_2012,Hirschinger_crosspolarization_2013}.
In such cases, it is convenient to use the estimation method based
on exploiting projective evolutions, which can be implemented with
magnetic field gradients, as shown in Ref. \citep{Alvarez2010,Xiao2006,alvarez_controlling_2012}.
Regardless of choosing the optimal way to estimate the coupling between
the spins, working in the QZE regime using projective measurements
allows us to estimate the coupling selectively without needing to
know the offset of the Hartmann-Hahn condition. This is a great advantage
in these complex systems.

\subsubsection{Three-spin couplings}

We now extend our focus to estimating the interacting couplings among
three spins. Specifically, we assume one spin $S$ and two spin $I$
again in the presence of a static field $B_{0}$ in the $z$ direction
and radiofrequency fields $B_{1,I}$ and $B_{1,S}$ in the $x$ direction.
The dipolar couplings between $S$ and both spins $I$ are denoted
as $b_{k}$, $k=1,2$, while the coupling between the spins $I$ is
denoted as $d$. For a three-spin system, as in an extended cross-polarization
experiment,$S$ represents for example a $^{13}C$ nucleus and the
two spins $I$, e.g. two protons $^{1}H$ \citep{chattah03,Hirschinger_crosspolarization_2011,Hirschinger_crosspolarization_2013,raya2017sensitivity,hirschinger2023analytical}.
The dipolar couplings between the proton $k$ and the carbon are 
\begin{align}
b_{k} & =-\frac{1}{2}\left(\frac{\mu_{0}\gamma_{I}\gamma_{S}}{4\pi r_{(k)}^{3}}\right)\frac{3r_{(k)z}^{2}-r_{(k)}^{2}}{r_{(k)}^{2}}, &  & k=1,2,
\end{align}
and the coupling between the protons is
\begin{align}
d & =-\frac{1}{2}\left(\frac{\mu_{0}\gamma_{I}^{2}}{4\pi r^{3}}\right)\frac{3r_{z}^{2}-r^{2}}{r^{2}}.
\end{align}

Similar to the two coupled spins case (Sec. \ref{subsec:Two-spin-coupling}),
the Hamiltonian in the double rotating frame preserves the total magnetization
of the system along the respective directions of the radiofrequency
fields. The Hamiltonian again acquires a block structure defined by
the total magnetization along $z$ of the full system $M=M_{I}+M_{S}$,
where $M_{I}=M_{1}+M_{2}$ is the total magnetization of the protons
along $z$ with $M_{1}$ and $M_{2}$ being the $z$-magnetization
of each proton, and $M_{S}$ is the $z$-magnetization of the $^{13}\mathrm{C}$.
The Hamiltonian induces transitions only between states of the form
$\{\ket{M_{I},\,M_{S}}\}$ and $\{\ket{M_{I}\pm1,M_{S}\mp1}\}$.

Symmetric and anti-symmetric states, $\ket{S}$ and $\ket{A}$, respectively,
are defined in the proton Zeeman basis as
\begin{align}
\ket{S} & =\frac{1}{\sqrt{2}}(\ket{+,-}+\ket{-,+}),\\
\ket{A} & =\frac{1}{\sqrt{2}}(\ket{+,-}-\ket{-,+}).
\end{align}
Within the total Zeeman subspace 
\[
\{\ket{S}\otimes\ket{+},\,\ket{+,+}\otimes\ket{-},\,\ket{A}\otimes\ket{+}\},
\]
the Hamiltonian block for $M=\frac{1}{2}$ is given by
\begin{align}
[H]_{M=\frac{1}{2}} & =\begin{pmatrix}\frac{1}{4}(\Sigma-\Delta)+\frac{1}{2}d & \frac{\sqrt{2}}{8}(b_{1}+b_{2}) & 0\\
\frac{\sqrt{2}}{8}(b_{1}+b_{2}) & (\frac{1}{4}\Sigma+\frac{3}{4}\Delta)-\frac{1}{4}d & \frac{\sqrt{2}}{8}(b_{2}-b_{1})\\
0 & \frac{\sqrt{2}}{8}(b_{2}-b_{1}) & \frac{1}{4}(\Sigma-\Delta)
\end{pmatrix}.
\end{align}
There are several realistic situations where the conditions $b_{1}=b_{2}=b$
or $b_{1}=-b_{2}=b$ hold, as it is the case determined by the symmetries
of the liquid crystal nCB molecules \citep{chattah03}. Therefore
for simplicity, we consider the case of $b_{1}=b_{2}$, where we see
that the transitions between $\ket{+,+}\otimes\ket{-}$ and $\ket{A}\otimes\ket{+}$
vanish. In this case, we obtain a polarization dynamics between only
two levels dictated by the effective Hamiltonian 
\begin{align}
\begin{pmatrix}\frac{1}{4}(\Sigma-\Delta)+\frac{1}{2}d & \frac{\sqrt{2}}{4}b\\
\frac{\sqrt{2}}{4}b & (\frac{1}{4}\Sigma+\frac{3}{4}\Delta)-\frac{1}{4}d
\end{pmatrix}.
\end{align}
Except for a constant component, this corresponds to a Hamiltonian
like the one considered in Eq. (\ref{eq:Hamiltonian}). Here, $\omega_{x}=-\frac{\sqrt{2}}{2}b$
depends on the dipolar coupling between the protons and the carbon,
and $\omega_{z}=\frac{1}{2}\Delta-\frac{3}{8}d$ depends on the offset
and the proton-proton interaction. For the case $b_{1}=-b_{2}$, exactly
the same result is obtained within the the Hamiltonian block $M=-\frac{1}{2}$
\citep{chattah03}. The proposed method can be implemented for estimating
the heteronuclear dipolar coupling $b$ when the interaction between
the protons is large, or if we do not know it and we want to estimate
the coupling $b$ regardless of the knowledge of $d$. Moreover, the
projective methods can also be useful for off-resonant polarization
transfers, as discussed in the previous Sec. \ref{subsec:Two-spin-coupling}.

As described in Eqs. (\ref{eq:quotient})-(\ref{eq:condition-low-polarization}),
for a large offset $\omega_{z}$, projective measurements become more
efficient for the estimation of $b$ compared with coherent evolutions
if the total available measurement time is bounded by $t<2\mathrm{e}\frac{\frac{1}{2}b^{2}+\left(\frac{1}{2}\Delta-\frac{3}{8}d\right)^{2}}{\frac{\sqrt{2}}{4}b^{3}}$.
For the case $b_{1}=-b_{2}$ a similar result is obtained.

\subsubsection{Generalization to many-spin systems}

The general properties for the comparison between estimation under
coherent and projective evolutions are not dependent of the details
of the quantum dynamics. Analogous to the case of two-level systems,
coherent and projective estimations are defined mainly by the second
and the first sum of Eq. (\ref{eq:QFI_lambda-1}) respectively. Coherent
estimation provides a QFI that, in general terms, evolves with the
square of the product between the evolution time and a factor dependent
on the parameter to be estimated (see Eq. (\ref{eq:QFI_coh_approx})).
Similarly, the QFI extracted from projective evolutions arises from
the exponential decay of a spin observable dynamics. Thus, optimal
evolutions are expected to be approximated by the characteristic decay
time induced by the stroboscopic measurements approach (see Eq. (\ref{eq:t_characteristic_stroboscopic})).
In general, the parameter values at which the QZE estimation approach
becomes advantageous over the coherent aproach, are when the transition
exchange probability between the relevant states is significantly
reduced. Yet, an important advantage of the QZE approach is that the
dependency of several parameters of the dynamics is further simplified
by the projective evolutions, as exploited in Ref. \citep{alvarez_controlling_2012}.
This simplification facilitates the estimation of couplings strengths
between spins, enabling the determination of the spin-spin coupling
network topology in many-body spin system.Coherent evolution is generally
very complex and difficult to use practically for determining the
entire coupling structure \citep{alvarez_controlling_2012,Luy1999,Tycko2007,Caravatti1983,Braunschweiler1983}.

Here demonstrate the implementation of a QZE estimation protocol as
a tool for inferring spin-spin couplings in many-spin systems. To
illustrate this phenomenon, we consider the Trotter-Suzuki expansion
to determine the quantum dynamics at short times, where the QZE approach
is manifested and becomes useful.

We consider a Hamiltonian with an isotropic interaction between the
spins as used in Ref. \citep{alvarez_controlling_2012} to showcase
the extension of our approach for estimating the couplings strengths
between spins in many-body systems. However, the results disussed
here are valid for Heisenberg-type interactions. The Hamiltonian is
\begin{align}
H & =\sum_{i}\omega_{i}^{z}I_{i}^{z}+\sum_{i<j}b_{ij}\mathbf{I}_{i}\cdot\mathbf{I}_{j},
\end{align}
where $i$ indexes the spins and $b_{ij}$ are the coupling strengths
between spins $i$ and $j$. The evolution operator can be expanded
using a first order Troter-Susuki expansion at short time to lead
\begin{align}
U(t) & =\mathrm{e}^{-iHt}\approx\prod_{i}\mathrm{e}^{-it\omega_{i}^{z}I_{i}^{z}}\prod_{i<j}\mathrm{e}^{-itb_{ij}\mathbf{I}_{i}\cdot\mathbf{I}_{j}}.
\end{align}
If the initial condition is given by $\rho(0)=1+\mu_{0}I_{i}^{z}$,
its quantum evolution will be determined by the evolution of the operator
$I_{i}^{z}$. At short times $t\ll b_{ij}^{-1}$, $(\omega_{i}^{z}-\omega_{j}^{z})^{-1}$
its evolution is given by 
\begin{align}
I_{i}^{z}\rightarrow I_{i}^{z}\left(1-\sum_{j\neq i}\frac{b_{ij}^{2}t^{2}}{8}\right)+\sum_{j\neq i}I_{j}^{z}\frac{b_{ij}^{2}t^{2}}{8}+\mathcal{O},
\end{align}
where $\mathcal{O}$ represents higher Trotter-Suzuki expansion orders
in time, and non/observable terms by monitoring the evolution of $I_{i}^{z}$
by non-demolition measurements \citep{alvarez_controlling_2012}.
The dynamic observed from the initially excited spin $\Tr\left[\rho(t)I_{i}^{z}\right]$,
in this short time regime, can be mapped to one given by a central
spin $i$ homogeneously coupled to the remaining spins $j$, with
an effective interaction $b=\sqrt{\frac{1}{N}\sum b_{ij}^{2}}$ where
$N$ is the number of spins \citep{eferraro10}. Then, the spins $j$
can be decimated to a single effective spin following the protocol
described in Ref. \citep{pastawski01}, thus allowing to reduce the
dynamics to one described by an effective two-spin system. Therefore,
within this quantum Zeno regime $\omega_{i}^{z}\tau,b_{ij}\tau\rightarrow0$,
$\alpha(t)\approx1-b^{2}t^{2}/8$ and the QFI is maximized at $\mathcal{F_{Q\mathrm{,max}}^{\mathrm{proj}}}\rightarrow32\varphi(\mu_{0})/b^{2}$
by measuring the spin-state $I_{i}^{z}$ at the total evolution time
$t_{\text{{max} }}\approx\frac{8\xi(\mu_{0})}{b^{2}\tau}$. This is
an important feature of this QZE estimation strategy, as it does not
require previous knowledge of the ofssets $\omega_{i}^{z}$ and the
couplings $b_{jl}$ with $j\ne l\ne i$ to make the inferrence efficient.

\section{Conclusions}

In summary, our study into the potential exploitation of the Quantum
Zeno Effect (QZE) to maximize information for quantum sensors represents
a step forward in quantum sensing technologies. Focusing on the general
features of the level avoided crossing (LAC) phenomenon in two-level
systems as a paradigm defining the Hamiltonian of the quantum sensor,
underscores the importance of the QZE in estimating the coupling strength—a
parameter essential for various quantum sensing applications.

We introduce the concept of information amplification by the QZE,
particularly in off-resonant conditions. Our findings reveal that
incoherent control, specifically through stroboscopic projective measurements,
may outperform coherent strategies for coupling strength estimation,
especially when facing time constraints due to decoherence. The use
of the quantum Fisher information as a metric for inference strategies
sheds light on the nuanced dynamics between coherent and incoherent
evolution in qubit-probe systems. Notably, our results indicate that,
under time constraints imposed by decoherence, the incoherent strategy
exhibits superior performance for large offsets.

We show practical applications supporting the advantages of the proposed
QZE inference protocol. We demonstrate its effectiveness in inferring
off-resonant AC magnetic fields and spin-spin couplings, offering
examples from two-spin to many-spin systems.

One of the key outcomes of our work is that achieving the quantum
Zeno regime enables selective inference of coupling strengths. This
strategy simplifies the qubit-probe dynamics and the inference procedure
by filtering out the complexity of the full system. For instance,
we demonstrate that in this regime, prior knowledge of the offset
or non-first neighbor spin-spin coupling to the sensor is not required.

The implementation of incoherent control, leveraging Quantum Non-Demolition
(QND) projective measurements and induced dephasing, emerges as a
versatile tool for steering qubit-probe evolution. The results of
QND measurements, crucial for guiding the system's dynamics, do not
necessitate readout, allowing emulation through various methods, such
as induced dephasing via random magnetic field gradients, $T_{2}$
relaxation, or stochastic interactions.

In essence, our work aims to contribute to the ongoing development
of quantum sensing methodologies, providing insights for optimizing
quantum sensor performance. By exploring incoherent control and strategically
choosing parameters, we hope our approach will open new possibilities
for enhancing quantum sensing capabilities across different applications.
Our findings, contribute to the collective efforts in precision measurement
techniques and lay the groundwork for potential advancements in quantum
technology.
\begin{acknowledgments}
This work was supported by CNEA; CONICET; ANPCyT-FONCyT PICT-2017-3156,
PICT-2017-3699, PICT-2018-4333, PICT-2021-GRF-TI-00134, PICT-2021-I-A-00070;
PIBAA 2022-2023 28720210100635CO, PIP-CONICET (11220170100486CO);
UNCUYO SIIP Tipo I 2019-C028, 2022-C002, 2022-C030; Instituto Balseiro;
Collaboration programs between the MINCyT (Argentina) and MOST (Israel).
B.R. acknowledges support from the Instituto Balseiro (CNEA-UNCUYO).
A. Z. and G.A.A. acknowledge support from CONICET.
\end{acknowledgments}

\appendix

\section{Quantum Fisher Information as a function of the polarization vector\label{sec:QFI_polarization}}

Starting from the QFI given by Eq. (\ref{eq:QFI_lambda-1}), we derive
the QFI in terms of the polarization vector (Eq. (\ref{eq:QFIfrompolars})).The
density matrix $\rho$, given by Eq. (\ref{eq:rho}), is diagonalized
by the vectors 
\begin{align}
\ket{\frac{1+\mathfrak{\mu}}{2}} & =\cos\left(\frac{\theta}{2}\right)\ket{\uparrow}+\mathrm{e}^{i\varphi}\sin\left(\frac{\theta}{2}\right)\ket{\downarrow},\\
\ket{\frac{1-\mathfrak{\mu}}{2}} & =\mathrm{e}^{-i\varphi}\sin\left(\frac{\theta}{2}\right)\ket{\uparrow}-\cos\left(\frac{\theta}{2}\right)\ket{\downarrow},\\
\end{align}
with eigenvalues $\frac{1}{2}\left(1+\mathfrak{\mu}\right)$ and $\frac{1}{2}\left(1-\mathfrak{\mu}\right)$
respectively, where $\mathfrak{\mu}$ is the magnitude, and $\theta$
and $\varphi$ are the azimuthal and polar angles of $\pmb{\mathfrak{\mu}}$.

The term related to the mixing of the QFI, Eq. (\ref{eq:QFI_lambda-1}),
takes the form: 
\begin{align}
\sum_{n:\lambda_{n}\neq0}\frac{(\partial_{\omega_{x}}\lambda_{n})^{2}}{\lambda_{n}} & =\frac{\left(\partial_{\omega_{x}}\left(\frac{1+\mathfrak{\mu}}{2}\right)\right)^{2}}{\frac{1+\mathfrak{\mu}}{2}}+\frac{\left(\partial_{\omega_{x}}\left(\frac{1-\mathfrak{\mu}}{2}\right)\right)^{2}}{\frac{1-\mathfrak{\mu}}{2}}\\
 & =\frac{1}{2}\left(\partial_{\omega_{x}}\mathfrak{\mu}\right)^{2}\left(\frac{1}{1+\mathfrak{\mu}}+\frac{1}{1-\mathfrak{\mu}}\right)\\
 & =\frac{\left(\partial_{\omega_{x}}\mathfrak{\mu}\right)^{2}}{1-\mathfrak{\mu}^{2}}=\frac{\left(\partial_{\omega_{x}}\pmb{\mathfrak{\mu}}_{r}\right)^{2}}{1-\mathfrak{\mu}^{2}}.
\end{align}

On the other hand, 
\begin{align}
\partial_{\omega_{x}}\ket{\frac{1+\mathfrak{\mu}}{2}} & =\partial_{\omega_{x}}\theta\,\frac{1}{2}\left(-\sin\left(\frac{\theta}{2}\right)\ket{\uparrow}+\mathrm{e}^{i\varphi}\cos\left(\frac{\theta}{2}\right)\ket{\downarrow}\right)\\
 & +i\partial_{\omega_{x}}\varphi\,\mathrm{e}^{i\varphi}\sin\left(\frac{\theta}{2}\right)\ket{\downarrow}\nonumber \\
 & =\mathrm{e}^{i\varphi}\left(-\frac{1}{2}\partial_{\omega_{x}}\theta\ket{\frac{1-\mathfrak{p}}{2}}+i\partial_{\omega_{x}}\varphi\,\sin\left(\frac{\theta}{2}\right)\ket{\downarrow}\right),\\
\partial_{\omega_{x}}\ket{\frac{1-\mathfrak{\mu}}{2}} & =\partial_{x}\theta\,\frac{1}{2}\left(\mathrm{e}^{-i\varphi}\cos\left(\frac{\theta}{2}\right)\ket{\uparrow}+\sin\left(\frac{\theta}{2}\right)\ket{\downarrow}\right)\\
 & -i\partial_{\omega_{x}}\varphi\,\mathrm{e}^{-i\varphi}\sin\left(\frac{\theta}{2}\right)\ket{\uparrow}\nonumber \\
 & =\mathrm{e}^{-i\varphi}\left(\frac{1}{2}\partial_{\omega_{x}}\theta\ket{\frac{1+\mathfrak{\mu}}{2}}-i\partial_{\omega_{x}}\varphi\,\sin\left(\frac{\theta}{2}\right)\ket{\downarrow}\right),\\
\end{align}
where (using $2\sin\left(\frac{\theta}{2}\right)\cos\left(\frac{\theta}{2}\right)=\sin(\theta)$),
\begin{align}
\abs{\mel{\frac{1\mp\mathfrak{\mu}}{2}}{\partial_{\omega_{x}}}{\frac{1\pm\mathfrak{\mu}}{2}}}^{2}=\frac{1}{4}\left[\left(\partial_{\omega_{x}}\theta\right)^{2}+\sin^{2}\theta\left(\partial_{\omega_{x}}\varphi\right)^{2}\right].
\end{align}

Considering also that
\begin{align}
\frac{\left(\frac{1\pm\mathfrak{\mu}}{2}-\frac{1\mp\mathfrak{\mu}}{2}\right)^{2}}{\frac{1\pm\mathfrak{\mu}}{2}+\frac{1\mp\mathfrak{\mu}}{2}} & =\mathfrak{\mu}^{2},
\end{align}
the second term takes of Eq. (\ref{eq:QFI_lambda-1}) the form
\begin{align}
2\sum_{n\neq m}\frac{(\lambda_{n}-\lambda_{m})^{2}}{\lambda_{n}+\lambda_{m}}\left|\left\langle \lambda_{m}\right|\partial_{\omega_{x}}\left|\lambda_{n}\right\rangle \right|^{2} & =\\
\mathfrak{\mu}^{2}\left[(\partial_{\omega_{x}}\theta)^{2}+\sin^{2}\theta(\partial_{\omega_{x}}\varphi)^{2}\right] & =\left(\partial_{\omega_{x}}\pmb{\mathfrak{\mu}}_{t}\right)^{2}.
\end{align}
 Finally, combining all the expressions, we arrive at Eq. (\ref{eq:QFIfrompolars})
\begin{align}
\mathcal{F_{Q}}(\omega_{x}) & =\frac{1}{1-\mathfrak{\mu}^{2}}\left(\partial_{\omega_{x}}\pmb{\mathfrak{\mu}}_{r}\right)^{2}+\left(\partial_{\omega_{x}}\pmb{\mathfrak{\mu}}_{t}\right)^{2}.
\end{align}

\section{Quantum Fisher Information (QFI) for projective measurements\label{ap1qfi}}

Considering Eqs. (\ref{eq:muvst}) and (\ref{eq:muz_proj}), we can
write the derivative as
\begin{align}
\partial_{\omega_{x}}\pmb{\mathfrak{\mu}}(n\tau+\Delta t)=\\
\mu_{0}\left[n(\alpha(\tau))^{n-1}\partial_{\omega_{x}}\alpha(\tau)\pmb\mu(\Delta t)+(\alpha(\tau))^{n}\partial_{\omega_{x}}\pmb\mu(\Delta t)\right].
\end{align}

Given that $\pmb\mu(t)$ is a unit vector for all $\omega_{x}$, $\pmb\mu\cdot\partial_{\omega_{x}}\pmb\mu=0$.
This implies that the squares of the radial and tangential components
are
\begin{align}
\left(\partial_{\omega_{x}}\pmb{\mathfrak{\mathfrak{\mu}}}_{r}\right)^{2} & =\mu_{0}^{2}n^{2}(\alpha(\tau))^{2(n-1)}(\partial_{\omega_{x}}\alpha(\tau))^{2},\\
\left(\partial_{\omega_{x}}\pmb{\mathfrak{\mathfrak{\mu}}}_{t}\right)^{2} & =\mu_{0}^{2}(\alpha(\tau))^{2n}\left(\partial_{\omega_{x}}\pmb\mu(\Delta t)\right)^{2}.
\end{align}
The QFI is is then given by
\begin{align}
\mathcal{F_{Q}}(\omega_{x}) & =\mu_{0}^{2}n^{2}\frac{(\alpha(\tau))^{2(n-1)}}{1-\mu_{0}^{2}(\alpha(\tau))^{2n}}\left(\partial_{\omega_{x}}\alpha(\tau)\right)^{2}\label{eq:Fq-proj-SI}\\
 & +\mu_{0}^{2}(\alpha(\tau))^{2n}\left(\partial_{\omega_{x}}\pmb\mu(\Delta t)\right)^{2},\nonumber 
\end{align}
where
\begin{align}
(\partial_{\omega_{x}}\alpha(\tau))^{2} & =\left(\frac{\omega_{x}}{\omega}\right)^{2}\left[2\frac{\omega_{z}^{2}}{\omega^{3}}(1-\cos(\omega\tau))+\frac{\omega_{x}^{2}}{\omega^{2}}\tau\sin(\omega\tau)\right]^{2},\\
\left(\partial_{\omega_{x}}\pmb\mu(\Delta t)\right)^{2} & =\frac{\omega_{x}^{4}}{\omega^{4}}\Delta t^{2}+2\frac{\omega_{x}^{2}\omega_{z}^{2}}{\omega^{5}}\sin(\omega\Delta t)\,\Delta t\\
 & +\frac{\omega_{z}^{2}}{\omega^{4}}\left[(1-\cos(\omega\Delta t))^{2}+\frac{\omega_{z}^{2}}{\omega^{2}}\sin^{2}(\omega\Delta t)\right].
\end{align}

The first term is constant as a function of the time between measurements
$\Delta t$, and the second term is bounded by $\mu_{0}^{2}(\alpha(\tau))^{2n}\max_{\Delta t\in\left[0,\,\tau\right)}\left(\partial_{\omega_{x}}\pmb\mu(\Delta t)\right)^{2}$.
At the projected measurement times $t=n\tau$, the first term of Eq.
(\ref{eq:Fq-proj-SI}) becomes Eq. (\ref{eq:QFI-proj}).

\bibliographystyle{apsrev4-1}
\phantomsection\addcontentsline{toc}{section}{\refname}\bibliography{bibliography_bak,mibib,biblioJMRO}

\end{document}